\documentclass[conference]{IEEEtran}
\usepackage{amsmath,amssymb,amsfonts}
\usepackage{graphicx}
\usepackage{textcomp}
\usepackage{algorithm}
\usepackage{amsmath}
\usepackage{xcolor}
\def\BibTeX{{\rm B\kern-.05em{\sc i\kern-.025em b}\kern-.08em
    T\kern-.1667em\lower.7ex\hbox{E}\kern-.125emX}}

\usepackage[utf8]{inputenc}
\usepackage[english]{babel}
\addto\captionsenglish{\renewcommand{\figurename}{Fig.}}
\usepackage[style=ieee]{biblatex}

\usepackage{comment}    
\usepackage[inline]{enumitem}   
\usepackage{amsmath}    
\usepackage{multirow}   
\usepackage{float}   
\usepackage{soul}   
\usepackage{csquotes}
\usepackage{hyperref}
\usepackage[T1]{fontenc}
\usepackage{float}

\usepackage{algpseudocode} 
\usepackage{subcaption} 
\usepackage{listings}
\lstdefinelanguage{VHDL}{
  morekeywords={
    library,use,all,entity,is,port,in,out,end,architecture,of,
    begin,and
  },
  morecomment=[l]--
}
\colorlet{keyword}{blue!100!black!80}
\colorlet{comment}{green!90!black!90}
\lstdefinestyle{vhdl}{
  language     = VHDL,
  basicstyle   = \ttfamily,
  keywordstyle = \color{keyword}\bfseries,
  commentstyle = \color{comment}
}

\makeatletter
\long\def\@makecaption#1#2{%
  \small
  \sbox\@tempboxa{#1. #2}%
  \ifdim \wd\@tempboxa >\hsize
    #1. #2\par
  \else
    \global \@minipagefalse
    \hb@xt@\hsize{\hfil\box\@tempboxa\hfil}%
  \fi}
\makeatother

\begin{document}

\title{FPGA Implementation of OTFS Modulation for 6G Communication Systems}

\author{Murat Isik\(^\dagger\), Malvin Nkomo, Anup Das, Kapil R. Dandekar
\thanks{\(^\dagger\)Corresponding Author.}
\thanks{All authors are with the Department of Electrical and Computer Engineering, Drexel University. This research is supported by the National Science Foundation under Grants CNS-1828236 and CNS-1816387. Any opinion, findings, and conclusion or recommendations expressed in this paper are those of the author(s) and do not necessarily reflect the reviews of the National Science Foundation.}
}

\IEEEoverridecommandlockouts

\maketitle

\IEEEpubidadjcol

\begin{abstract}
Sixth-generation (6G) communication systems are poised to accommodate high data-rate wireless communication services in highly dynamic channels, with applications including high-speed trains, unmanned aerial vehicles, and intelligent transportation systems. Orthogonal frequency-division multiplexing (OFDM) modulation suffers from performance degradation in such high-mobility applications due to high Doppler spread in the channel. The recently proposed Orthogonal Time Frequency Space (OTFS) modulation scheme outperforms OFDM in terms of supporting a higher transmitter (Tx) and receiver (Rx) user velocity. Additionally, the highly-dynamic time-frequency (TF) channel has little effect on OTFS modulated signals, which enables the realization of low-complexity pre-processing architectures for implementing massive-multiple input multiple outputs (MIMO) based OTFS systems. However, while OTFS has received attention in the literature from a theory and simulation perspective, there has been comparatively little work on real-time FPGA implementation of OTFS waveforms.  Thus, in this paper, we first present a mathematical overview of OTFS modulation and then describe an FPGA implementation of OTFS implementation on hardware. Power, area, and timing analysis of the implemented design on a Zynq UltraScale+ RFSoC FPGA are provided for benchmarking purposes.
\end{abstract}

\section{Introduction}

Multipath propagation channels are doubly selective due to time dispersion (frequency selectivity) and Doppler shift (time selectivity)~\cite{yucek2008time}. Both of these phenomena are due to the highly varying nature of the time-frequency (TF) channel. Fourth-generation (4G) and fifth-generation (5G) communication systems address the time-dispersive nature of the channel using orthogonal frequency division multiplexing (OFDM), which mitigates inter-symbol interference (ISI) through the use of a cyclic prefix and longer symbol duration \cite{ozan2020zero,kuti2021evaluation}. However, a high-velocity communication scenario produces high Doppler shifts, which destroys the orthogonality of the sub-carriers generated by the OFDM modulator, thus causing inter-carrier interference (ICI) \cite{wang2006performance}. While OFDM cannot mitigate the effects of high Doppler shifts, the recently proposed orthogonal time frequency space (OTFS) modulation scheme is highly effective in these environments due to its delay-doppler characteristic\cite{hadani2018otfs}.

OTFS modulation multiplexes the information in the Delay-Doppler (DD) domain in contrast to the TF domain used in OFDM modulation. This is advantageous as the DD domain representation of any rapidly-varying TF channel is both slowly varying and sparse in nature \cite{ramachandran2020otfs}. 
Consequently, modulating in the DD domain allows the received signal to interact with a DD channel that exhibits both sparsity and slow variability. This interaction improves the performance in terms of throughput and error rate, compared to a signal modulated in the TF domain that must deal with a rapidly changing TF channel. OTFS modulation has shown significant improvement in performance compared to OFDM modulation in high-Doppler scenarios where the user velocity was as high as 500 km/h in the 4 GHz band \cite{hadani2018otfs}. Considering mmWave communication, OTFS outperformed OFDM in the 28 GHz band at a user velocity of 40km/h \cite{hadani2017orthogonal,surabhi2019otfs}. However, these results were obtained using software simulation alone. 

The ability of Field Programmable Gate Arrays (FPGAs) to implement complex digital signal processing algorithms in real-time is making them increasingly popular in wireless communication systems. FPGAs are superior to traditional digital signal processors (DSPs) in wireless communication applications due to their high parallelism, flexibility, power efficiency, and speed. 
A complex modulation scheme like OFDM requires real-time processing of large amounts of data. The high parallelism and low latency of FPGAs make them ideal for processing OFDM signals. 
Several processing blocks are required for OFDM, including the fast-Fourier transform (FFT), inverse fast-Fourier transform (IFFT), modulation, demodulation, and channel estimation. Various wireless communication systems can be easily prototyped by modifying IP cores used by FPGAs to implement these processing blocks. 
FPGAs provide real-time processing of data, which is essential for wireless communication systems to provide reliable and efficient communication \cite{desai2021efficient, cardarilli2021design, mohamed2011novel, kadiran2005design}. All of these characteristics also make FPGAs ideal for prototyping OTFS systems prior to standardization.

We propose that FPGA implementation would be a powerful solution for evaluating wireless communication modulation schemes such as OTFS. The high parallelism, high flexibility, low power consumption, and high processing speed of FPGAs make them an ideal platform for the processing of large amounts of data required by wireless communication systems. The use of customizable IP cores in FPGA-based implementations of OTFS provides a high degree of flexibility and allows the designer to optimize the design for specific system requirements, and to prototype different variations of the waveform to help inform the design of future 6G standards.

\section{Related Works}
This section summarizes recent literature on OTFS modulation.  Hadani et al. \cite{hadani2017orthogonal} have analyzed the performance of OTFS modulation in millimeter wave systems under the influence of phase noise, Doppler spread, and delay spread defined by various cellular band standards.
Emanuele et al. \cite{raviteja2019effective} analyzed the diversity of OTFS over two-path channels. The results show that OTFS has a lower bit-error rate (BER) than OFDM in a number of varying case scenarios.  Chockalingam et al. \cite{murali2018otfs} described a low-complexity detection scheme according to Markov chain Monte Carlo (MCMC) sampling techniques and a Pseudo-Noise (PN) graph sequence-based channel estimate technique for the DD domain. 


There is comparatively little research on real-time FPGA hardware implementations of OTFS waveforms.   \cite{thaj2019otfs,marsalek2020otfs} have performed the Software-defined radio (SDR) implementation of an OTFS modem. \cite{thaj2019otfs} studied the performance of OTFS and OFDM modulation systems in real indoor wireless channel scenarios. \cite{marsalek2020otfs} presents the performance of an OTFS system, where the received signal extracted from a 60GHz millimeter wave carrier frequency is passed through a Linear Minimum Mean Square Error (LMMSE) equalizer.  

In a study by \cite{shadangi2023vlsi}, a novel VLSI architecture for OTFS modulation was proposed for high-speed vehicular communication scenarios. The authors presented a report on resource utilization for implementation of OTFS on an FPGA board, as well as a demonstration of the input-output relationship of an OTFS signal on a single input, single output channel under additive white gaussian noise conditions. A CORDIC processor was used for non-linear function generation ($\sin(\theta)$, $\cos(\theta)$, etc) for designing the transmitter and receiver. This design was implemented on a Xilinx Zynq-7 FPGA board, thus demonstrating a power-efficient approach. Our paper uniquely provides a detailed performance analysis, including power, area, and timing, which were previously unexplored in hardware implementations of OTFS, thus contributing to the advancement of practical applications for 6G communications systems.

More recently \cite{dora2023low}, a low-complexity implementation of an OTFS transmitter was proposed using a fully parallel and pipelined hardware architecture. FFTs and IFFTs were parallel and depth pipelined on an FPGA to accelerate OTFS execution, resulting in high accuracy and performance. The authors also proposed an optimized OTFS transmitter architecture with a modified Booth multiplier and memory, which needed fewer hardware resources while providing higher performance. The OTFS hardware architecture achieved a bandwidth of 196.67 Tbps at 139.64 MHz maximum operating frequency, thus making it suitable for future 5G and 6G wireless communications standards. Furthermore, the optimized hardware architecture reduced the LUTs on the Virtex-7 FPGA board by around 20\% compared to a conventional OTFS transmitter.

To the best of our knowledge, and contrary to the previous works on OTFS which mostly focused on theoretical explanation and software simulation, this paper is one of the first to present OTFS implementation on an FPGA board.

\section{OTFS Modulation}

\begin{figure}
    \centering
    \includegraphics[width=0.485\textwidth]{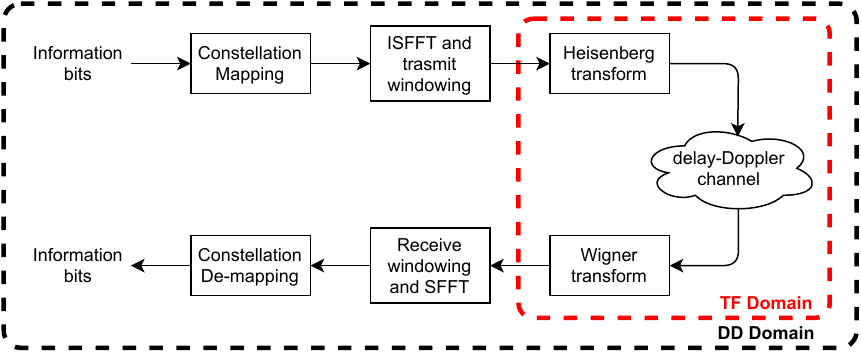}
    \caption{Block diagram of an OTFS transceiver system}
    \label{fig_1_OTFS_transceiver}
\end{figure}

OTFS modulation can be implemented as an extension to the existing OFDM modulation framework for 4G communication systems \cite{ramachandran2020otfs}. An OTFS transceiver system starts with the constellation mapping of the information bits on a discretized DD plane. The data symbols in the DD domain are then converted to the TF domain symbols using the two-dimensional (2D) Inverse Symplectic Finite Fourier transform (ISFFT) at the transmitter side. This is followed by the Heisenberg transform, which converts the TF symbols to a time-domain signal. This signal is then pulse-shaped with a suitable window and is sent over the channel. The transmitted OTFS frame also includes additional pilot symbols for channel estimation. In the receiver, the received time-domain signal is converted back to the symbols in the DD domain with the help of the Wigner transform, followed by an SFFT operation on the received signal. Fig. \ref{fig_1_OTFS_transceiver} provides a block diagram of the OTFS system.

The mathematical representations of the above process are shown below:
\begin{itemize}

    \item Discretized DD and TF grids are represented as:
    \[\Omega_{DD} = \{(p \Delta\nu,q \Delta\tau), p=0,...,N-1;q=0,...,M-1 \}\]
    \[\Sigma_{TF} = \{(nT,m \Delta f), n=0,...,N-1;m=0,...,M-1 \}\]
    where, \[\left \{ \Delta\nu = \frac{1}{NT},~\Delta\tau = \frac{1}{M \Delta f} \right \}\]
    The discretized grids are shown in Fig. \ref{grid}.
    
    \begin{figure}
		\centering
		\includegraphics[width=0.485\textwidth]{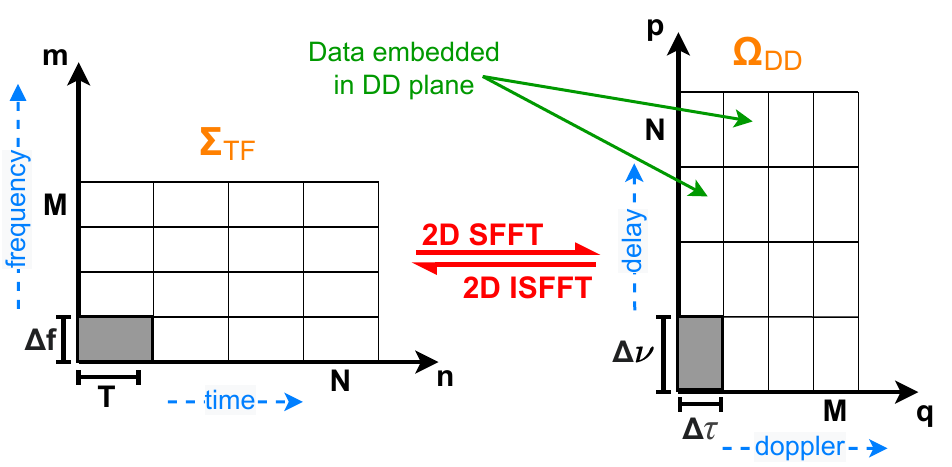}
		\caption{TF and DD plane}
		\label{grid}
    \end{figure}
	    
    \item The input bit stream is converted to its respective information symbols in the IQ plane and then mapped to the 2D DD grid \(\Omega_{DD}\). The resultant data frame is represented by \(x[p,q]\), having a dimension of \(N\times M\). \(x[p,q]\) resides in the DD domain.
	
	\item Next, the ISFFT operation in conjunction with a transmit windowing function maps the data frame in the DD domain to its equivalent TF domain \(\Sigma_{TF}\), represented by \(X_\rho[n,m]\):
    \[X_\rho[n,m] = \frac{1}{\sqrt{NM}} \sum_{p=0}^{N-1} \sum_{q=0}^{M-1} x_\rho[p,q]e^{j2 \pi \left( \frac{np}{N}- \frac{mq}{M} \right)}\]
    \[X[n,m] = W_{tx}[n,m]\cdot X_\rho[n,m]\]
    where, \(W_{tx}[n,m]\) is the square summable transmit windowing function.
    
    \item In the final step, the TF data frame is converted to a continuous time waveform for transmission using the Heisenberg transform:
    \[s(t) = \sum_{n=0}^{N-1} \sum_{m=0}^{M-1} X[n,m]e^{j2\pi m\Delta f(t-nT)} g_{tx}(t-nT)\]
    where \(g_{tx}\) is the transmit pulse.
    
\end{itemize}

\section{FPGA Implementation}

\begin{figure}
    \centering
    \includegraphics[width=0.46\textwidth]{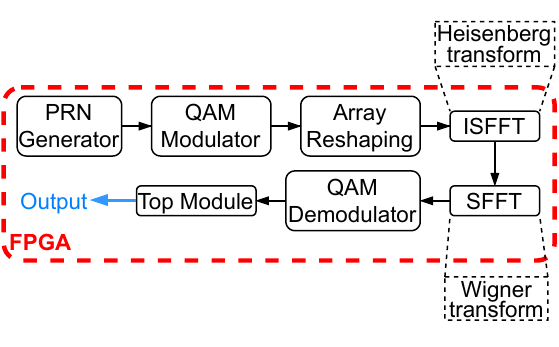}
    \caption{Top Hardware Architecture}
    \label{fig_2_FPGA_block}
\end{figure}

The architecture design implements the various processes in parallel, thus consuming less processing time. Xilinx Intellectual Property (IP) cores and custom modules are used in the implementation that is shown in Figure \ref{fig_2_FPGA_block}. Modules are defined using VHDL based on the AXI Interface. The modules in the design use \(12\)-bit data configuration for the real and imaginary parts. A Xilinx FFT core is used to improve the efficiency and speed of the FFT and IFFT blocks used inside the ISFFT module, which has a maximum instantaneous frequency of \(250\) MHz and can operate up to \(50\) frames/sec.  This architecture has six parts, described in the following sub-sections:
\begin{itemize}
    \item Random-Bit Generator
    \item QAM Modulator
    \item Array reshaping
    \item ISFFT and Heisenberg
    \item Wigner and SFFT
    \item QAM Demodulator
\end{itemize}

\subsection{Random-Bit Generator}
\begin{figure}[h]
    \centering
    \includegraphics[width=0.4\textwidth]{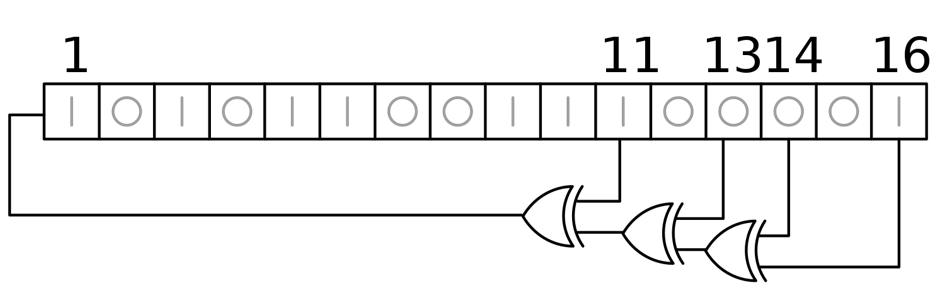}
    \caption{LFSR Structure}
    \label{lfsr}
\end{figure}

The generation of random bits is a fundamental requirement for testing various communication and information processing systems. In this regard, a simple and efficient technique for generating pseudo-random binary sequences (PRBS) is the use of a linear feedback shift register (LFSR). In this current work, a 16-bit LFSR is utilized to generate a \(8192\) bit long PRBS sequence. The feedback loop for the LFSR is implemented using a simple configuration, as shown in Fig. \ref{lfsr}. The LFSR is configured as a 16-bit shift register, and the feedback polynomial is defined as follows:

\begin{equation}
f(x) = 1 + x^{11} + x^{13} + x^{14} + x^{16}
\end{equation}

The LFSR is initialized with all \(1\)'s, and the output bit is generated by XORing the first, third, fourth, and sixth bits of the shift register. The generated bit is output and the shift register is updated for each clock cycle.

The code also includes control logic to set the number of random bits to be generated based on the desired modulation order. The number of bits is computed as follows:
\begin{itemize}
    \item For \(M=4\), \(8192\) bits are generated \((4096 \times 2)\)
    \item For \(M=8\), \(12288\) bits are generated \((4096 \times 3)\)
    \item For \(M=16\), \(16384\) bits are generated \((4096 \times 4)\)
    \item For \(M=32\), \(20480\) bits are generated \((4096 \times 5)\)
\end{itemize}

\begin{figure}[h]
    \centering
    \includegraphics[width=0.4\textwidth]{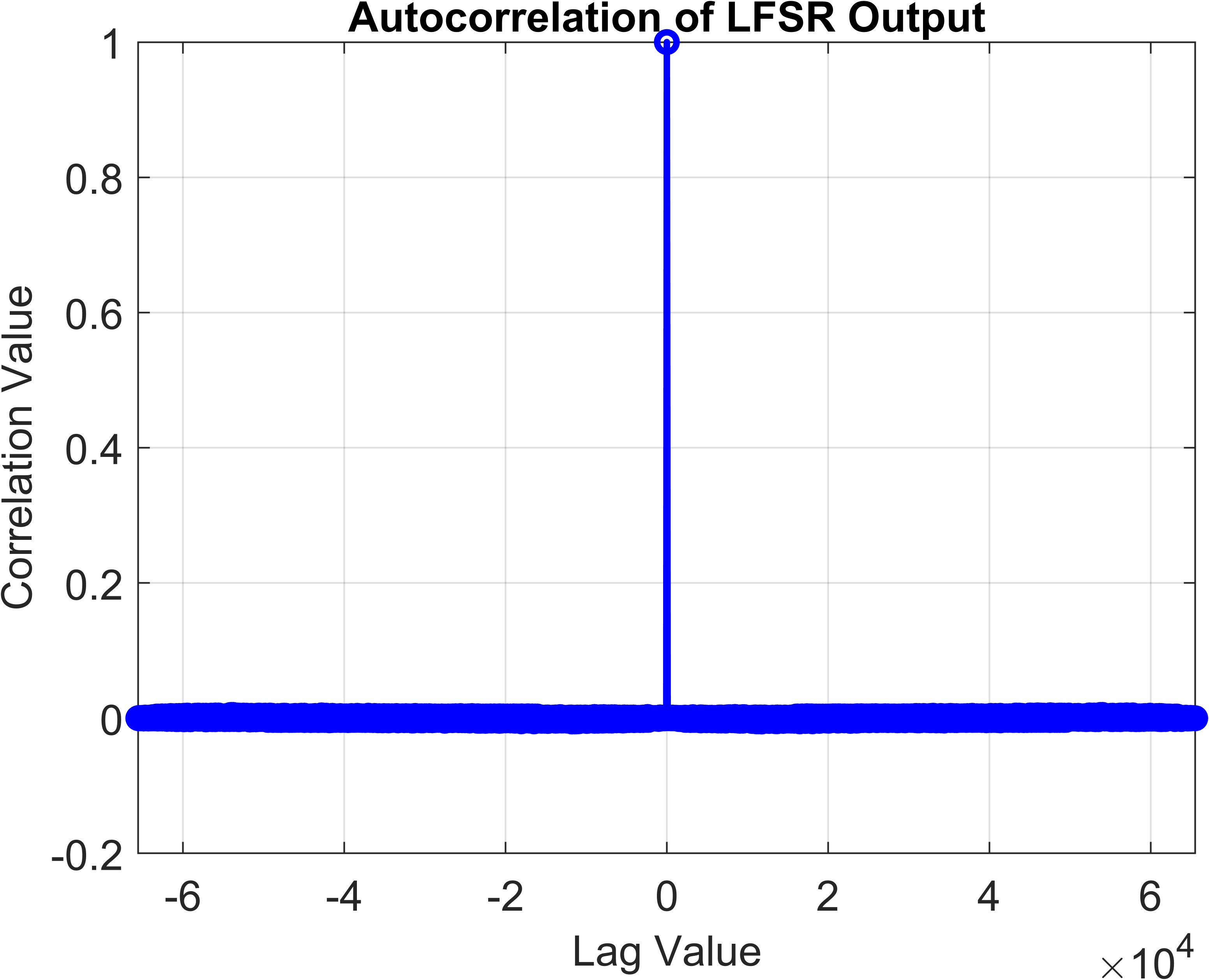}
    \caption{Autocorrelation.}
    \label{Autocorrelation}
\end{figure}

After generating the required number of random bits, the control logic sets the generator to idle mode. The generated random bits were analyzed for their auto-correlation function, which is depicted in Fig. \ref{Autocorrelation}. 
This figure illustrates the degree of correlation between consecutive bits. A lower auto-correlation implies a sequence with characteristics that resemble randomness. It is worth noting, however, that while the 16-bit LFSR-generated sequence meets the requirements of this study, its predictability characteristics may vary depending on the application context.

The generated random bit sequence is evaluated in both MATLAB and VHDL environments. The auto-correlation analysis of the \(16\)-bit LFSR generated bit sequence demonstrates that it closely resembles white noise, indicating that the generated sequence is statistically random. Moreover, a VHDL testbench is designed to compare the performance of the PRBS generator implemented in VHDL with that of the PRBS generated in MATLAB. It is observed that the generated pseudo-random bit sequences in both environments are virtually identical.

LFSR has been widely adopted for generating pseudorandom bit sequences due to its simplicity and efficiency \cite{payne1973pseudorandom, stkepien2012application}. We have chosen a \(16\) LFSR-based PRBS generator to create statistically random bit sequences in this work. These sequences are essential for a wide range of practical applications, such as in communication and information processing systems. This choice is supported by the inherent properties of LFSRs, which provide good statistical properties and implementation efficiency \cite{panda2012fpga}. Our VHDL implementation of the generator has demonstrated reliability and performance consistent with expectations, further validating its utility. However, we acknowledge that different scenarios might warrant the use of other methods. For instance, Xilinx offers various on-chip random bit generators that could be leveraged depending on the specific requirements of the system \cite{Xilinx2012}. Therefore, while the \(16\) LFSR-based PRBS generator serves our purposes in this study, we recognize the existence and potential suitability of other well-established solutions in diverse scenarios.


\subsection{QAM Modulator and Array Reshaping}
A 4-quadrature amplitude modulation (4-QAM) scheme is implemented using VHDL. The input to the 4-QAM modulation consists of \(20480\) bits, which are used to generate \(4096\) complex numbers. The generated complex numbers are then reshaped into a \(64\times 64\)-sized array for further processing. It has been observed that the \(4096\) complex numbers produced by the VHDL implementation are identical to those produced by MATLAB's built-in (\texttt{qammod}) function when tested using a testbench. The proposed implementation supports multiple modulation schemes, including \(4\)-QAM, \(8\)-QAM, \(16\)-QAM, and \(32\)-QAM. The length of the input bits into the QAM modulator is defined by the modulation scheme. If the modulation order is \(N\), the input data is generated using the formula \(4096\times log_2(N)\). The symbol values are defined as per the look-up tables (LUTs) used for the QAM modulation. The QAM modulator generates the output by directly accessing the LUTs and assigning the values to the real and imaginary parts of the modulation output.

The constellations generated by MATLAB satisfy the unit average power property, whereby the sum of the squares of the magnitudes of its complex symbols equals one. The constellation's complex symbols are represented by floating-point numbers, with the real and imaginary parts represented as separate values, and floating-point values are converted  to fixed-point notation, with a 10-bit resolution, thus resulting in 12-bit signed numbers with \(2\)'s complement signed \(2.10\) format. For instance, the floating-point value for \(971/1024\) is represented as \(0.948\) in decimal notation.



\subsection{Inverse Symplectic Finite Fourier transform (ISFFT) and Heisenberg Transform}

In this module, a \(64\times 64\) matrix containing \(4096\) complex numbers are fed into an IFFT block and then is  followed by a transpose operation. If the transpose operation is omitted, the system resembles the classic OFDM scenario. To complete the ISFFT operation, this data stream is finally fed into an FFT block.

The obtained IFFT results are stored in a Xilinx Block Random Access Memory (BRAM) with a capacity of \(64\times 64 = 4096\) values. The first column of the \(64\times 64\) IFFT result is stored consecutively at addresses 0 to 63, the second column at addresses 64 to 127, and so on. To obtain the transpose of the IFFT output, the saved data from the BRAM is then processed to feed the output of the module to the transposed matrix. Specifically, the first column of the transposed IFFT matrix corresponds to the values stored at addresses 0, 64, 128,.., 4096-64, and the second column corresponds to the values stored at addresses 1, 65, 129,..., 4096-63, and so on. A snippet of the main portion of the VHDL code is shown in Alg. \ref{VHDL_1}.
 
\begin{algorithm}[h]
\begin{lstlisting}[style=vhdl]
when RECORD_IFFT_DATA =>
RecState <= RECORD_IFFT_DATA;
addrb <= (others => '0');
if(m_axis_data_tvalid = '1') then
wea <= "1";
dina <= m_axis_data_tdata(42 downto 27)& 
m_axis_data_tdata(18 downto 3);
addra <= std_logic_vector(unsigned(addra)+1);
if (unsigned(addra) = 4094) then
RecState <= OUTPUT_OTFS_DATA_0;
end if;
end if;
\end{lstlisting}
\caption{VHDL code for recording data from the Xilinx IFFT IP core.}
\label{VHDL_1}
\end{algorithm}

Due to the limitations of the Xilinx FFT IP core, it is not possible to feed the entire matrix into the FFT core. To address this limitation, we propose using the \texttt{Feed\_FFT\_Data} state, where the columns of the matrix are fed to the FFT IP core one by one. The incoming data is fed to the FFT IP core operating at 100 \(MHz\), and the resulting FFT data is stored in Xilinx BRAM. A snippet of the main portion of the VHDL code is shown in Alg. \ref{VHDL_2}.

\begin{algorithm}[h]
\begin{lstlisting}[style=vhdl]
when FEED_FFT_DATA =>
state <= FEED_FFT_DATA;
if(QAMDataValid = '1') then
s_axis_data_tvalid <= '1';
s_axis_data_tdata <= std_logic_vector(resize(signed(QAMDataIm),16)) &
std_logic_vector(resize(signed(QAMDataRe),16));
DataCount <= std_logic_vector(unsigned(DataCount) + 1);
if (DataCount(5 downto 0) = "111111") then
s_axis_data_tlast <= '1';
end if;
if (unsigned(DataCount) = 4095) then
State <= IDLE;
end if;
end if;
\end{lstlisting}
\caption{VHDL code for feeding data to the Xilinx FFT IP core.}
\label{VHDL_2}
\end{algorithm}

\subsection{Wigner Transform and Symplectic Finite Fourier transform (SFFT)}

The Wigner transform is an essential component in the OTFS demodulation process. At the receiver, the Wigner Transform and SFFT are applied to demodulate the signal back to the DD domain. Our approach involves breaking down the Wigner transform into distinct operations, such as complex multiplication, summation, and Fourier transformations. This enables us to focus on optimizing each operation independently before integrating them into the complete transformation. By designing customized IP cores for these operations, we were able to optimize the performance of the Wigner transform on the FPGA. The Wigner Transform is then performed by computing the FFT of the reshaped matrix and dividing it by the square root of the total number of subcarriers.

After generating the FFT IP core, we complete the Wigner transform process by performing complex multiplication and conjugation of the input signals before feeding them into the FFT IP core, thus utilizing available Xilinx IP cores. Finally, the design is synthesized and implemented on the hardware, and the results are then verified through simulations in MATLAB and on-hardware tests.  The accuracy of the results is also compared with that obtained from MATLAB simulation results.

The SFFT is an essential component in the OTFS modulation technique. The algorithm begins by reshaping the input array into a matrix with dimensions corresponding to the number of subcarriers and the number of symbols. To implement the SFFT component, the output from the Wigner Transform is transposed the inverse FFT of the transposed matrix is computed. The result is normalized by division by the square root of the number of symbols times the number of subcarriers. We use the IFFT Xilinx IP core to calculate the IFFT of the FFT of the transposed matrix Y, which amounts to the SFFT of the transposed matrix Y. Finally, we divide the IFFT output by the square root of the number of symbols divided by the number of subcarriers to obtain the demodulated OTFS signal. By using the IFFT Xilinx IP core, we can efficiently delegate computationally intensive tasks, such as FFT and IFFT calculations, to dedicated hardware, thus improving both the performance and power efficiency of the design.

\subsection{QAM Demodulator}

In the final stage of the OTFS design, the received signal must be demapped from the complex symbols back to the original information bits. This is achieved by identifying the closest constellation point to the received symbol and then demapping it back to the corresponding bit sequence. The demapping process takes into account the fixed-point notation used earlier and converts the complex symbols back to the floating-point representation.

\section{Performance Analysis}
\label{sec_ımp}

A summary of the resource utilization report for the Zynq UltraScale+ XCZU28DR FPGA can be found in Table \ref{table:utilization}. It contains a variety of resources including LUTs, LUTRAMs, FFs, IOs, and BUFGs. The ``Utilization" column shows how many resources were used in the design, while the ``Available" column shows how many resources are available in the FPGA. A percentage of utilization is also computed for reference. Based on the table, the utilization percentage for LUT is 0.40\%, LUTRAM is 0.27\%, FF is 0.41\%, IO is 39.19\%, and DSP is 0.94\%. There is enough capacity for more clock buffers based on the lower utilization rate for BUFG.

\begin{table}[h]
\centering
\caption{Resource utilization summary}
\label{table:utilization}
\begin{tabular}{|c|c|c|c|}
\hline
\multicolumn{4}{|c|}{Zynq UltraScale+ XCZU28DR} \\ \hline
\textbf{Resource} & \textbf{Utilization} & \textbf{Available} & \textbf{\% Utilization} \\ \hline
LUT & 1708 & 425280 & 0.40 \\
LUTRAM & 586 & 213600 & 0.27 \\
FF & 3446 & 850560 & 0.41 \\
IO & 136 & 347 & 39.19 \\
DSP & 40 & 4272 & 0.94 \\ \hline
\end{tabular}
\end{table}

\begin{figure}[h]
    \centering
    \includegraphics[width = 0.4\textwidth]{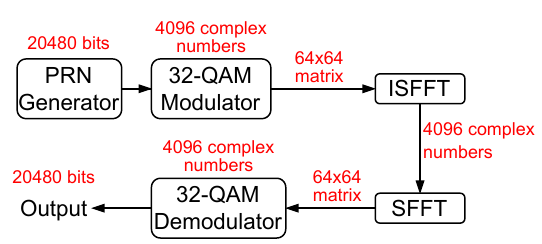}
    \caption{Data sizes and the dimension of output data streams.}
    \label{data_dim}
\end{figure}

Table \ref{table:table2} provides an overview of the system's capabilities and performance characteristics, and Fig. \ref{data_dim} lists the size and dimension of the output streams from each block. A \(32\)-QAM modulation scheme was used for the final design. OTFS modulation uses these four symbols (from \(32\)-QAM's mapping table) to map information from the DD-domain to the TF-domain. Two bits of information are contained in each symbol. Multiplying the number of symbols by the number of bits per symbol results in \(20480\) bits, which denotes the total number of bits from the LFSR. The FFT blocks sampling rate is listed as \(61.44\) MHz. The system's performance can be improved by using a higher sampling rate. This system consumes \(1.45\) Watt of power. There is a latency of \(12.17 \; \mu\)s listed for the system. This latency is quite negligible for any real-time implications. This design lists a throughput of \(503.31\) Gbits/sec. The throughput of a wireless channel is the speed at which data can be transmitted. This metric takes into account the system's power consumption and throughput and provides a measure of how efficiently the system could transmit data wirelessly. In this case, the system is able to transmit 155.38 Gigabits of data per second per watt of power consumed. In comparison to recent OTFS transmitter implemented by \cite{dora2023low}, which achieved a bandwidth of 196.67 Tbps at a maximum operating frequency of 139.64 MHz, our design has demonstrated a throughput of 503.31 Gbits/sec at a 400 MHz operation frequency. Moreover, our system's power consumption of 1.45 W is significantly lower than the average power consumed by existing systems. This higher efficiency potentially makes our system more suitable for future 6G wireless communication standards, in terms of both speed and power consumption. Considering the stringent requirements of emerging 6G standards, such as higher data rates, lower latencies, and increased energy efficiency, our system demonstrates significant potential. However, further advancements in ultra-reliable low-latency communications and machine-type communication scalability would be needed to fully meet these future demands.

\begin{table}[h]
\centering
\caption{General OTFS System Implementation Parameters.}
\label{table:table2}
\begin{tabular}{|c|c|}
\hline
Parameters & Specifications \\ \hline
Modulation scheme & \(32\)-QAM \\ \hline
Operation Frequency & \(400\) MHz \\ \hline
Total Bits & \(20480\) bits \\ \hline
FFT Sampling Rate & \(61.44\) MHz \\ \hline
Power & \(1.45\) W \\ \hline
Latency &  \(12.17\) \(\mu\)s \\ \hline
Throughput & \(503.31\) Gbits/sec \\ \hline
Throughput Efficiency & \(347.11\) Gbits/sec/W \\ \hline
\end{tabular}
\end{table}

\begin{figure*}[h]
    \centering
    \includegraphics[width = 0.90\textwidth]{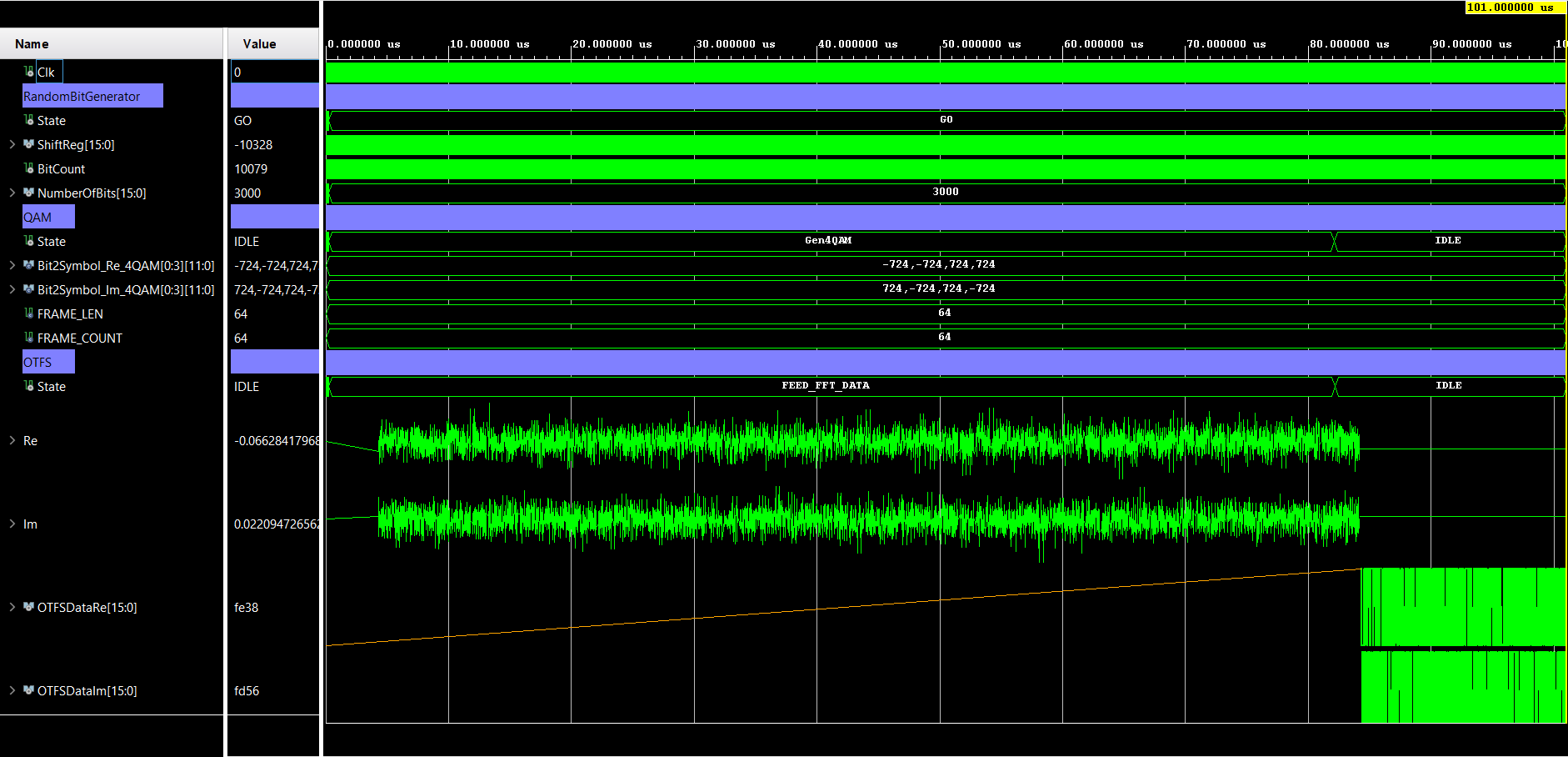}
    \caption{Simulation result of the system}
\end{figure*}

\begin{figure}[h]
    \centering
    \includegraphics[width = 0.48\textwidth]{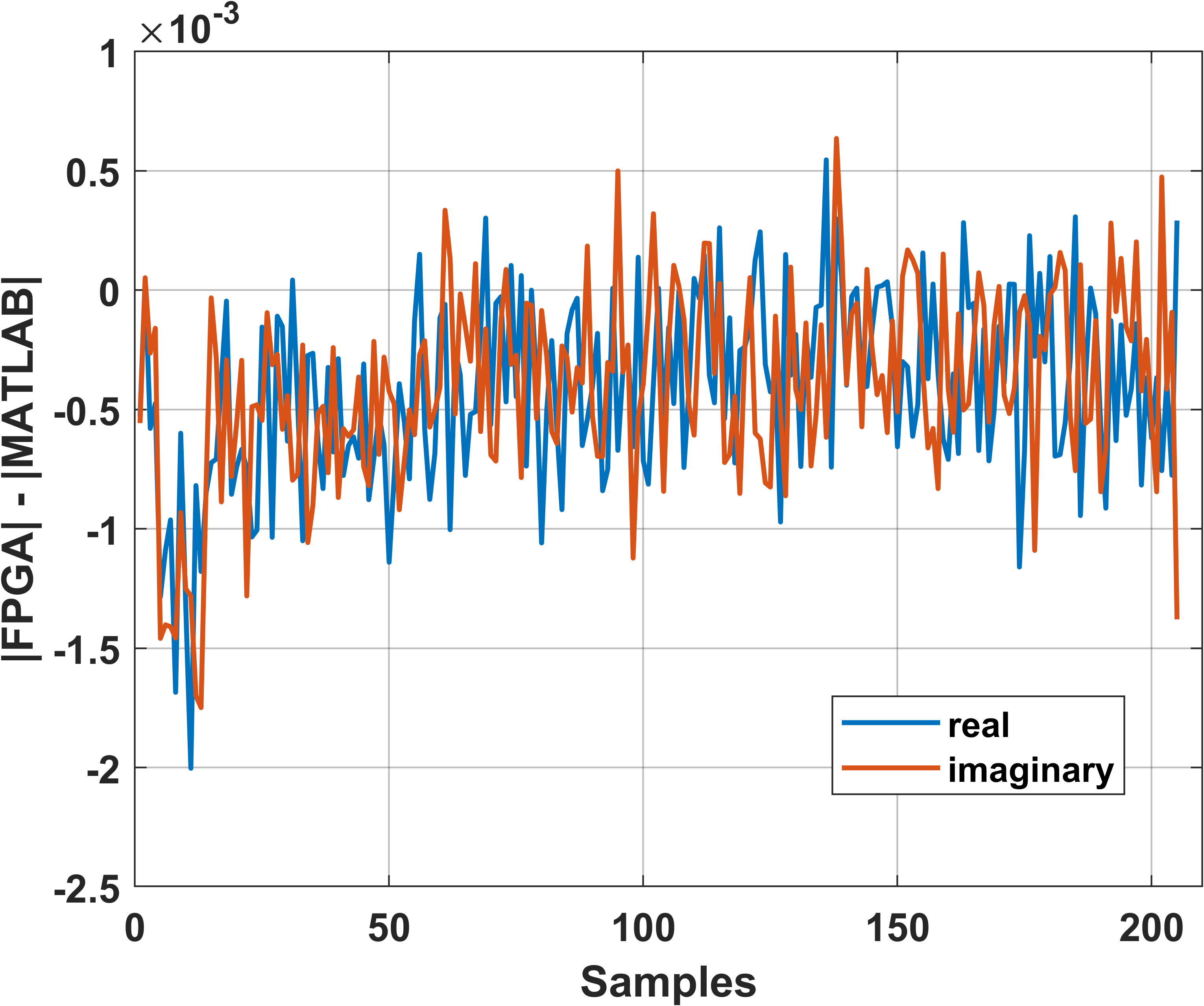}
    \caption{Error difference between VHDL and MATLAB implementation of the modulated signal.}
\end{figure}

\begin{figure}[h]
    \centering
    \includegraphics[width = 0.48\textwidth]{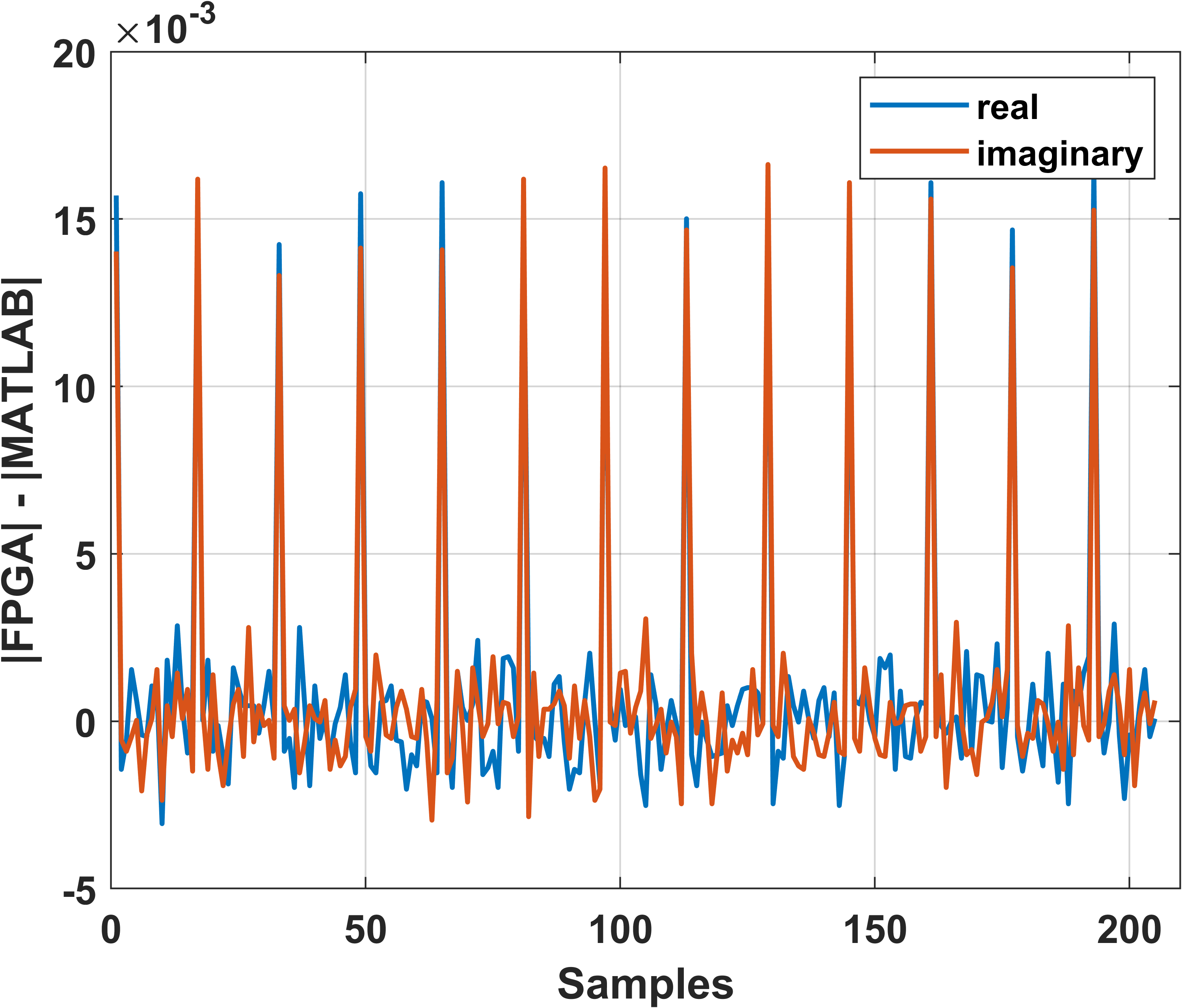}
    \caption{Error difference between VHDL and MATLAB implementation of the demodulated signal.}
\end{figure}

     
    

\section{Conclusions}

This paper describes the implementation of an OTFS  system on a FPGA. The resilience of OTFS to high Doppler shifts makes it attractive for high user velocity scenarios. Also, because of the sparse and slow varying nature of the DD channel, the channel has less impact on the DD multiplexed data frame. This work presents a computationally efficient design of an OTFS modulation system on a Zynq UltraScale+ RFSoC FPGA. 
Resource requirements are reported to demonstrate implementation efficiency. As we delve deeper into the practicalities of OTFS, it becomes evident that understanding its behavior under diverse environmental conditions is crucial. Moreover, while our design is efficient, there's always room for further optimization, potentially leading to even better performance metrics. Challenges, both foreseen and unforeseen, await as we transition from controlled environments to real-world deployments. Addressing these challenges, which could range from interference management to integration complexities, will be instrumental in the broader adoption of OTFS. Future work includes the integration of an RF frontend on a NI Ettus USRP X410, which leverages the same FPGA used in this paper, to enable over-the-air (OTA) transmission and reception for studying the effectiveness of OTFS in realistic wireless scenarios.

\printbibliography

@ARTICLE{wang2006performance,
  author={Wang, T. and others},
  journal={IEEE Trans. Wirel. Commun.}, 
  title={Performance Degradation of {OFDM} Systems Due to Doppler Spreading}, 
  year={2006},
  volume={5},
  number={6},
  pages={1422--1432},
}

@ARTICLE{ozan2020zero,
  author={Ozan, W. and others},
  journal={IEEE Commun. Lett.}, 
  title={Zero Padding or Cyclic Prefix: Evaluation for Non-Orthogonal Signals}, 
  year={2020},
  volume={24},
  number={3},
  pages={690--694},
}

@INPROCEEDINGS{kuti2021evaluation,
  author={Kuti, A. Y. and Abdelkareem, A. E.},
  booktitle={2021 IEEE Int. Conf. Commun., Netw. and Satellite (COMNETSAT)}, 
  title={Evaluation of Low-Density Parity-Check Code with {16-QAM OFDM} in a Time-Varying Channel}, 
  year={2021},
  pages={128--134},
}

@ARTICLE{hadani2018otfs,
  author={Hadani, R. and Monk, A.},
  journal={arXiv preprint arXiv:1802.02623}, 
  title={{OTFS}: A New Generation of Modulation Addressing the Challenges of {5G}}, 
  year={2018},
}

@article{ramachandran2020otfs,
  title={{OTFS}: A new modulation scheme for high-mobility use cases},
  author={Ramachandran, M. K. and Surabhi, G. D. and Chockalingam, A.},
  journal={Journal of the Indian Institute of Science},
  volume={100},
  number={2},
  pages={315--336},
  year={2020},
  publisher={Springer}
}

@inproceedings{surabhi2019otfs,
  title={{OTFS} modulation with phase noise in {mmWave} communications},
  author={Surabhi, G. D. and Ramachandran, M. K. and Chockalingam, A.},
  booktitle={2019 IEEE 89th Vehicular Technology Conference (VTC2019-Spring)},
  pages={1--5},
  year={2019},
  organization={IEEE}
}

@inproceedings{thaj2019otfs,
  title={{OTFS} modem {SDR} implementation and experimental study of receiver impairment effects},
  author={Thaj, T. and Viterbo, E.},
  booktitle={2019 IEEE International Conference on Communications Workshops (ICC Workshops)},
  pages={1--6},
  year={2019},
  organization={IEEE}
}

@inproceedings{marsalek2020otfs,
  title={{OTFS} modulation and influence of wideband {RF} impairments measured on a 60 {GHz} testbed},
  author={Marsalek, R. and Blumenstein, J. and Sch{\"u}tzenh{\"o}fer, D. and Pospisil, M.},
  booktitle={2020 IEEE 21st International Workshop on Signal Processing Advances in Wireless Communications (SPAWC)},
  pages={1--5},
  year={2020},
  organization={IEEE}
}

@inproceedings{desai2021efficient,
  title={Efficient implementation technique for {OFDM} on {FPGA}},
  author={Desai, A. and Gupta, A. and Jambhale, M. and Chavan, V.},
  booktitle={Proceedings of the 4th International Conference on Advances in Science \& Technology (ICAST2021)},
  year={2021}
}

@inproceedings{cardarilli2021design,
  title={Design and {FPGA} implementation of a low power {OFDM} transmitter for narrow-band {IoT}},
  author={Cardarilli, G. C. and Di Nunzio, L. and Fazzolari, R. and La Cesa, R. and Re, M.},
  booktitle={CEUR Workshop Proceedings},
  volume={3092},
  pages={60--65},
  year={2021}
}

@article{mohamed2011novel,
  title={A novel implementation of {OFDM} using {FPGA}},
  author={Mohamed, M. A. and Samarah, A. S. and Allah, M. I. F.},
  journal={International Journal of Computer Science and Network Security},
  volume={11},
  number={11},
  pages={43--48},
  year={2011}
}

@article{kadiran2005design,
  title={Design and implementation of {OFDM} transmitter and receiver on {FPGA} hardware},
  author={Kadiran, K. A. B.},
  journal={University Technology Malaysia, Electrical Engineering},
  year={2005}
}

@article{shadangi2023vlsi,
  title={{VLSI} architecture for implementing {OTFS}},
  author={Shadangi, A. R. and Das, S. S. and Chakrabarti, I.},
  year={2023}
}

@article{dora2023low,
  title={Low complexity implementation of {OTFS} transmitter using fully parallel and pipelined hardware architecture},
  author={Dora, S. K. and Mishra, H. B. and Sahoo, M.},
  journal={Journal of Signal Processing Systems},
  pages={1--10},
  year={2023},
  publisher={Springer}
}

@INPROCEEDINGS{hadani2017orthogonal,
  author={Hadani, R. et al.},
  booktitle={2017 IEEE MTT-S International Microwave Symposium (IMS)}, 
  title={Orthogonal Time Frequency Space ({OTFS}) Modulation for Millimeter-Wave Communications Systems}, 
  year={2017},
  pages={681--683},
}

@ARTICLE{yucek2008time,
  author={Yucek, T. and Arslan, H.},
  journal={IEEE Trans. Veh. Technol.}, 
  title={Time Dispersion and Delay Spread Estimation for Adaptive {OFDM} Systems}, 
  year={2008},
  volume={57},
  number={3},
  pages={1715--1722},
}

@article{raviteja2019effective,
  title={Effective diversity of {OTFS} modulation},
  author={Raviteja, Patchava and Hong, Yi and Viterbo, Emanuele and Biglieri, Ezio},
  journal={IEEE wireless communications letters},
  volume={9},
  number={2},
  pages={249--253},
  year={2019},
  publisher={IEEE}
}

@inproceedings{murali2018otfs,
  title={On {OTFS} modulation for high-Doppler fading channels},
  author={Murali, K Rꎬ and Chockalingam, A},
  booktitle={2018 Information Theory and Applications Workshop (ITA)},
  pages={1--10},
  year={2018},
  organization={IEEE}
}

@article{payne1973pseudorandom,
  title={Pseudorandom numbers for mini-and microcomputers: A generalized feedback shift register algorithm},
  author={Payne, WH},
  journal={Behavior Research Methods \& Instrumentation},
  volume={5},
  number={2},
  pages={93--98},
  year={1973},
  publisher={Springer}
}

@inproceedings{stkepien2012application,
  title={Application of the {DLFSR} generators in spread spectrum communication},
  author={St{\k{e}}pie{\'n}, Rafa{\l} and Walczak, Janusz},
  booktitle={Proceedings of the 19th International Conference Mixed Design of Integrated Circuits and Systems-MIXDES 2012},
  pages={555--558},
  year={2012},
  organization={IEEE}
}

@manual{Xilinx2012,
  title = {System Generator for {DSP} Reference Guide},
  author = {{Xilinx, Inc.}},
  year = {2012},
  edition = {14},
  month = oct,
}

@inproceedings{panda2012fpga,
  title={FPGA implementation of 8, 16 and 32 bit {LFSR} with maximum length feedback polynomial using {VHDL}},
  author={Panda, Amit Kumar and Rajput, Praveena and Shukla, Bhawna},
  booktitle={2012 International Conference on Communication Systems and Network Technologies},
  pages={769--773},
  year={2012},
  organization={IEEE}
}
\end{document}